\documentclass[3p]{elsarticle_mod}

\usepackage{lineno,hyperref}

\journal{Computer Physics Communications, vol. 237, p. 42-46 (2019)}

\begin{document}

\begin{frontmatter}

\title{Compact representation of one-particle wavefunctions and scalar fields obtained from electronic-structure calculations}


\author[ad1,ad2,ad3]{Sergey V. Levchenko}
\cortext[corrauth]{Corresponding author}
\ead{levchenko@fhi-berlin.mpg.de}

\author[ad2]{Matthias Scheffler}

\address[ad1]{Skolkovo Institute of Science and Technology, Skolkovo Innovation Center, 3 Nobel Street, 143026 Moscow, Russia}
\address[ad2]{Fritz-Haber-Institut der Max-Planck-Gesellschaft, Faradayweg 4-6, 14195 Berlin, Germany}
\address[ad3]{National University of Science and Technology ``MISIS'', 119049 Moscow, Russia}

\begin{abstract}
We present a code-independent compact representation of one-electron wavefunctions and other volumetric data (electron density, electrostatic potential, etc.) produced by electronic-structure calculations. The compactness of the representation insures minimization of digital storage requirements for the computational data, while the code-independence makes the data ready for ``big data'' analytics. Our approach allows to minimize differences between original and the new representation, and is in principle information-lossless. The procedure for obtaining the wavefunction representation is closely related to construction of natural atomic orbitals, and benefits from the localization of Wannier functions. Thus, our approach fits perfectly any infrastructure providing a code-independent tool set for electronic-structure data analysis.
\end{abstract}


\end{frontmatter}


\section{Introduction}
Self-consistent one-electron wavefunctions (WFs) contain detailed information on the electronic structure (ES) of a system in the mean-field approximation. A quick access to WFs would allow for an exhaustive analysis of many ES properties. WFs can be either calculated on demand or stored on disk (by disk we mean any device for storing information). Nowadays, some ES calculations are acceptably accurate and fast, so that storage of WFs can be avoided. For example, local-density and generalized gradient approximations to density-functional theory (DFT) allow calculating systems with tens of atoms on a single CPU in seconds. However, larger systems and more advanced ES methods (e.g., DFT with hybrid functionals or many-body perturbation theory) would require much more time and/or computational resources, which would inhibit ES data analysis across materials space. Thus, storage of WFs and other derived data such as electronic density or Hartree potential is highly desirable.

For a material model with periodic boundary conditions, the wavefunctions $\psi_{i\bf k}^{\sigma}({\bf r})$ are expressed in terms of basis functions $\phi_{\alpha}({\bf r})$:
\begin{equation}
  \psi_{i\bf k}^{\sigma}({\bf r})=\sum_{\alpha} C_{i\bf k}^{\sigma \alpha}\phi_{\alpha}({\bf r})\, ,
  \label{eq:basisrep}
\end{equation}
where $i$ is the band index, ${\bf k}$ is the k-point, and $\sigma$ is the spin. The same expression can be used for non-periodic systems by limiting the k-point index to ${\bf k} = {\bf 0}$.

Electronic-structure theory uses several, very different forms of $\phi_{\alpha}({\bf r})$ in its numerous computational implementations. These are for example (i) Gaussian-type orbitals (GTOs), (ii) plane waves (PWs), (iii) numeric atomic orbitals (NAOs), (iv) combination of the above, and more (including real-space grid-based codes such as Octopus~\cite{andrade2015real} and Parsec~\cite{kronik2006parsec}, and wavelet-based codes such as BigDFT~\cite{mohr2014daubechies}). Each of these basis-set types have their advantages and disadvantages. The absence of a single standard basis-set type presents a challenge for a code-independent representation of the WF data. Such a representation is needed to make the data generated by different ES codes directly comparable and interpretable by a machine (artificial intelligence) for data analytics, e.g., for finding outliers and hidden correlations within the data and with other theoretical or/and experimental data. Moreover, basis sets optimal for storage or for ES calculations usually have different requirements. For example, atomic orbitals form a most compact basis set for an atom, but Gaussian functions are more computationally efficient due to analytic integral evaluations.

Our goal is to find an optimal universal basis set to store WF data. We are particularly interested in {\em all-electron} calculations, which give access to important properties such as electric field gradients and nuclear magnetic resonance (NMR) shifts. In contrast, PW calculations are typically combined with pseudopotentials, and therefore the valence WFs near the atomic nuclei are incorrect, i.e., pseudoized. Nevertheless, approximate WFs with all the nodes can be straightforwardly restored from the pseudopotential calculations~\cite{van1993first} using the same formalism as the projector augmented-wave method (PAW)~\cite{blochl1994projector}. And the core electrons in a pseudopotential calculation are frozen to those used in in the creation of the pseudopotential.

\section{Formalism}
The expansion coefficients $C_{i\bf k}^{\sigma \alpha}$ and the functions $\phi_{\alpha}({\bf r})$ contain complete information for describing WFs at any point in space. Depending on the functional form, $\phi_{\alpha}({\bf r})$ can be stored as either a small set of parameters (k-point for each PW, or contraction coefficients and exponents for each GTO) for analytic functional forms, or calculated on demand (as NAOs) from a set of potentials represented by parameterized analytic forms (e.g., Coulomb potential in atomic Schr{\"o}dinger equation). In overwhelming majority of calculations localized basis functions (GTOs and NAOs) depend only on atomic species, and are otherwise system-independent. This is practical for ES calculations, because it is not clear how to efficiently adapt basis sets on-the-fly during self-consistent-field (SCF) cycle. However, for storage purposes this strategy may not be the optimal one, because in this case ES is already known, and there is no need to have a larger basis set for the sake of transferability. Moreover, storage puts more stringent requirements on transferability, because a WF representation that is compact in one basis may not be as compact in another basis.

In view of the above, we argue that NAOs are most flexible and therefore best suited for code-independent WF storage format. The numerical representation gives much flexibility to the form of the basis functions. With this flexibility, we can pursue the following strategy for obtaining the most compact representation of wavefunctions and derived scalar fields.

{\em Atom-centered numerical grids.} Strategies for constructing efficient and accurate atom-centered grids for NAOs are well established~\cite{delley1990all,treutler1995efficient}. Each NAO is expressed as a product of a radial and a spherical harmonic function. The radial part is represented as a spline interpolation on a radial grid. Following~\cite{baker1994effect}, we suggest to use logarithmic radial grids described by the following formula:
\begin{equation}
  r(s)=r_{\rm outer}\frac{\log(1-[s/(N_{\rm r}+1)])}{\log(1-[N_{\rm r}/(N_{\rm r}+1)])}\, ,
\end{equation}
where $r_{\rm outer}$ is the radius of the outer-most shell, $N_{\rm r}$ is the number of grid points, and $s$ = $\{1,\ldots , N_{\rm r}\}$ ($r(0)$ = 0 corresponds to the atomic center). From our extensive experience with NAO-based electronic-structure calculations with the FHI-aims package~\cite{blum2009ab} we conclude that $r_{\rm outer}$ = 7-8~\AA~is sufficient for a very accurate representation of WFs. If necessary, a higher accuracy can be achieved by simply placing additional shells at integer fractions of the original grid, e.g., $s$ = 1/2, 3/2, $\ldots$, 2$N_{\rm r}$ + 1/2. This allows us to specify a small base grid to reduce the storage requirements when possible, and easily interconvert between different grid densities for data analysis. The optimal base grid $N_{\rm r}$ values for each atom are tabulated in the species default settings of the FHI-aims package~\cite{blum2009ab}. Depending on the atom, the values vary between 24 and 80. The accuracy and transferability of the grids constructed in the way described above have been extensively tested by applications of the DMol$^3$~\cite{delley2000molecules} and FHI-aims packages to a large variety of systems, including molecules, clusters, and periodic systems of dimensions 1 to 3 (e.g., crystalline solids, nanowires, surfaces, and layered materials). 
In non-NAO-based approaches atom-centered grids are often used to represent WFs within a sphere centered at the nucleus, to reduce the number of PW basis functions. In all cases, we suggest to use code-independent grids.

In case of delocalized basis sets such as plane waves, or any non-atom-centered basis sets, an atom-centered basis set has to be first generated such that it accurately represents the set of one-electron states to be stored. For a numerical representation on a finite grid a {\em complete} set of possible radial functions is determined by the number of grid points and the type of interpolation between the points. Various NAO-based codes successfully use spline interpolation. Here we propose to apply basis splines (B-splines)~\cite{curry1966polya,de1976splines} for the radial parts of atomic functions. The angular parts are spherical harmonics truncated at a certain maximum angular momentum $l_{max}$. The benefit of B-splines is that they are localized along the radial coordinate, so that it is easy to select only B-splines that overlap with the regions where the non-atom-centered basis functions are applied. For example, in the linearized augmented plane-wave (LAPW) method there is a cutoff radius around each nucleus. Within the cutoff radius basis functions are atom-centered, while outside the basis functions are plane waves. Therefore, only B-splines that are non-zero outside the cutoff radius are needed to represent the PW part of the LAPW basis set. The completeness of the atom-centered representation of the selected electronic states can be quantified by a spillage parameter~\cite{sanchez1995projection,sanchez1996analysis}.

{\em Finding the minimal NAO basis set to represent the WFs to be stored with a desired accuracy.} For this purpose, the flexibility of NAOs exhibits its key advantages. To find the minimal set of NAOs, we employ a procedure similar to that of finding natural atomic orbitals~\cite{reed1985natural,reed1988intermolecular}. Although originally formulated for finite systems, an extension to periodic models is almost straightforward~\cite{dunnington2012generalization}. The procedure is based on diagonalization of the density matrix:
  \begin{equation}
    \Gamma ({\bf r}, {\bf r'})=\sum_{\sigma} \sum_{i{\bf k}}f_{i{\bf k}}^{\sigma}\psi^{\sigma *}_{i{\bf k}}({\bf r})\psi^{\sigma}_{i{\bf k}}({\bf r'})\, ,
  \end{equation}
  where $f_{i{\bf k}}^{\sigma}$ are occupation numbers. The method determines a compact set of linear combinations of basis functions with maximum overlap with the space spanned by the occupied states. Our task is similar, but our aim is to find a compact basis set for representation of states to be stored, independent of their actual occupation. The resulting NAOs are termed subspace-optimized atomic orbitals (SOAOs). Thus, $\Gamma ({\bf r}, {\bf r'})$ should be interpreted differently: $f_{i{\bf k}}^{\sigma}$ are non-zero only for states that we aim to store. In most practical cases all non-zero $f_{i{\bf k}}^{\sigma}$ should be identical (e.g., all set to 1). The radial parts of SOAOs are stored on the atom-centered grids, considering a spline interpolation along the radial coordinate (see below the discussion on storage memory requirements).

  Construction of natural AOs and consequently SOAOs is a multi-step procedure~\cite{reed1985natural,reed1988intermolecular}. It starts from calculation of pre-orthogonalized SOAOs by diagonalizing $(A,l,m)$ blocks of the density matrix, where $A$ denotes an atom, and $(l,m)$ are the angular momentum quantum numbers. Averaging of the $(l,m)$ blocks of the ``density matrix'' over the $2l+1$ values of $m$ for each atom and $l$ is performed to preserve spherical symmetry. As a result, we obtain a minimal set of NAOs (pre-SOAOs) that have maximum overlap with the set of WFs to be stored. The contribution of each pre-SOAO to the set is quantified by the eigenvalues of the modified density matrix (in the case of natural orbitals, the eigenvalues would be occupations). If all pre-SOAOs were orthogonal to each other, the eigenvectors corresponding to non-vanishing eigenvalues would constitute the minimal basis for accurate representation of the WFs. However, due to non-zero overlaps with pre-SOAOs on neighboring atoms, the eigenvalues of diffuse pre-SOAOs will not decay. Similar to natural AOs, we suggest to apply the occupancy-weighted symmetric orthogonalization (OWSO)~\cite{carlson1957orthogonalization,reed1985natural,reed1988intermolecular} procedure to pre-SOAOs, where instead of occupancy the eigenvalues of the atomic angular momentum blocks of the modified density matrix are used. OWSO yields orthogonal orbitals that preserve the shape of AOs with higher occupations as much as possible. Although OWSO yields orbitals which do not have spherical symmetry and are linear combinations of pre-SOAOs located on different atoms, they still can be assigned to particular atoms based on the location of the pre-SOAO with the highest contribution to a given post-OWSO orbital. As the next step, the modified density matrix is re-diagonalized in the basis of the post-OWSO orbitals on each atom, and the resulting orbitals with non-negligible eigenvalues (above a threshold) are selected. If we denote radial parts of pre-SOAOs on atom $A$ for angular momentum $l$ as $\chi_{\beta l}^A(r)$, the orbitals after the re-diagonalization are expressed as:
  \begin{equation}
    \phi_{\alpha}({\bf r})=\sum_A \sum_{lm} \left(\sum_{\beta} C_{\beta lm}^A \chi_{\beta l}^A(r)Y_{lm}(\Omega _A) \right)\, ,
    \label{eq:OWSO}
  \end{equation}
  where $C_{\beta lm}^A$ are obtained from the OWSO procedure, and $Y_{lm}(\Omega _A)$ are spherical harmonics centered at atom $A$. The radial functions $\sum_{\beta}C_{\beta lm}^A \chi_{\beta l}^A(r)$ should be collected on each atom $A$ for each $(l,m)$ channel, normalized, and linearly independent radial functions should be selected by diagonalizing their overlap matrix and removing the eigenvectors with small eigenvalues. The remaining radial functions within each $l$ channel are then orthogonalized to give the final set of atom-centered functions.

  \section{Discussion}
    Following the above steps, we obtain the minimal set of NAO basis functions that accurately represent the WFs to be stored. The flexibility of the NAO functional form allows to reduce the number of basis functions to a convenient minimal size, and, consequently, also the number of expansion coefficients in equation~\ref{eq:basisrep}. This flexibility comes with a small storage overhead because the radial parts of basis NAOs have to be stored on the atom-centered radial grids. If the number of basis functions is $N_{basis}$ and the number of WFs to be stored is $N_{states}$, then the total number of expansion coefficients is $N_{states}N_{basis}$ (assuming there is no sparsity). In case of the SOAOs, the number of functions $N_{SOAO} < N_{basis}$, but they require additional $N_{SOAO}N_{r}$ numbers for storage, where $N_{r}$ is the average number of radial grid points per atom (in practice $N_r \sim 150$). As the number of WFs to be stored for a given material is increased, $N_{SOAO}$ approaches $N_{states}$, and the overhead tends to $N_{states}N_{r}$.
  This storage overhead is not large ($\sim$1~kB per WF), and it can be further reduced by finding a parameterized set of atom-centered potentials such that a minimal number of solutions of Schr{\"o}dinger equations for these potentials is complete in the space spanned by SOAOs. Below we describe an approach for finding such potentials.

  The problem can be formulated in terms of minimization of the following functional for each atom in the system:
  \begin{equation}
    F[\{V_{\alpha}\}]=\sum_{\alpha l}\int_0^{\infty} \left[ \sum_{\beta} \left(-\frac{1}{2}U^{(l)}_{\alpha\beta}\frac{d^2\eta_{\beta l}}{dr^2} + \left( V_{\alpha}(r)+\frac{l(l+1)}{2r^2} \right)U^{(l)}_{\alpha\beta}\eta_{\beta l}-E_{\alpha l}U^{(l)}_{\alpha\beta}\eta_{\beta l}  \right) \right]^2r^2dr
    \label{eq:APF}
  \end{equation}
  with respect to unknown spherical potentials $V_{\alpha}(r)$, the eigenvalues $E_{\alpha l}$ for each angular momentum $l$, and unitary matrices $U^{(l)}_{\alpha\beta}$ under the condition that we obtain as small number of different $V_{\alpha}(r)$ as possible, so that we reduce storage requirements for the potentials. The summands in the square brackets of Eq.~\ref{eq:APF} are Schr\"{o}dinger equations for an arbitrary unitary transformation of radial parts $\eta_{\beta l}$ of the SOAOs within each $l$-channel. The unitary transformation is introduced to make use of the fact that we are only interested in restoring SOAOs up to a unitary transformation, which gives us additional flexibility for reducing the number of different parameters in the set of $V_{\alpha}$ functions. Minimizing $F[\{V_{\alpha}\}]$ simply means maximizing the overlap between the Hilbert spaces spanned by $N_{\eta}$ eigenvectors of potentials $V_{\alpha}$ and the radial functions $\eta_{\beta l}$, with $N_{\eta}$ being the total number of $\eta_{\beta l}$ functions. The multiplier $r^2$ outside the square brackets in Eq.~\ref{eq:APF} is needed to insure stability of the numeric integration on a logarithmic grid. The global minimum $F[\{V_{\alpha}\}]=0$ can be always reached if all $V_{\alpha}$ are allowed to be different. Indeed, the solution is obtained by inverting the Schr\"{o}dinger equation for each orbital, similar to the first step in constructing a pseudopotential.

  Using the definition of a unitary matrix, $({\bf U}^{(l)})^{\rm T}{\bf U}^{(l)}={\bf U}^{(l)}({\bf U}^{(l)})^{\rm T}={\bf I}$, and the orthonormality of $\eta_{\beta l}$ within each $l$-channel, it can be easily shown that $F[\{V_{\alpha}\}]$ is a positive definite (or semidefinite) quadratic polynomial with respect to elements of symmetric matrices $C_{\beta\beta '}^l$ = $\sum_{\alpha}U_{\alpha \beta}^{(l)}E_{\alpha l}U_{\alpha \beta '}^{(l)}$ and parameters of the potentials, provided $V_{\alpha}$ are parameterized linearly, i.e., $V_{\alpha}(r)$ = $\sum_i v_i^{(\alpha)}f_i(r)$, where $f_i(r)$ are some known functions (see below). Therefore, we can perform convex minimization of $F[\{V_{\alpha}\}]$ with respect to $v_i^{(\alpha)}$ and $C_{\beta\beta '}^l$ using compressed sensing to reduce the number of {\em different} potentials $V_{\alpha}$ for a given accuracy threshold. For this task, clustered least absolute shrinkage and selection operator (LASSO)~\cite{she2010sparse} is employed. LASSO converts the combinatorially hard problem of finding the most sparse solution (i.e., the solution with the minimal number of parameters) of a linear regression to a convex optimization problem. This is achieved by adding an $l_1$ penalty (the sum of absolute values of parameters) to the minimized function. In clustered LASSO, the $l_1$ penalty is a sum of absolute {\em differences} between the parameters rather than the parameters themselves. In our case, the optimized spherical potentials $V_{\alpha}^{\rm opt}$ for a given atom are obtained as:
    \begin{equation}
      V_{\alpha}^{\rm opt} = {\rm arg\,min}_{\{V_{\alpha}\},\{{\bf C}^l\}}\left( F[\{V_{\alpha}\}] + \lambda ||\Delta v||_1 \right)\, ,
      \label{eq:vopt}
    \end{equation}
    where
     \begin{equation}
       ||\Delta v||_1 = \sum_i\sum_{\alpha < \beta}|v_i^{(\alpha)}-v_i^{(\beta)}|\, .
       \label{eq:norm}
    \end{equation}
     In this approach, differences in individual parameters in $V_{\alpha}$ are penalized, so that a sparse solution is obtained, but not necessarily with a minimal number of different $V_{\alpha}$ (because two potentials are different as long as at least one parameter is different). If $V_{\alpha}$ are stored as a vector, e.g., ${\bf v}^{(1)}$ = $(v_1^{(1)}, v_2^{(1)}, \ldots)$, plus the matrix $v_i^{(\alpha)}-v_i^{(1)}$, $\alpha > 1$, the latter will be sparse, but the indices of non-zero elements also need to be stored. Alternatively, we can minimize the number of different $V_{\alpha}$, rather than the number of different $v_i^{(\alpha)}$. In this case, only the different ${\bf v}^{(\alpha)}$ vectors need to be stored, without any indices.

     The number of different $V_{\alpha}$ can be minimized using the following stepwise LASSO algorithm. First, Eq.~\ref{eq:vopt} is used with the penalty term from Eq.~\ref{eq:norm}. If $F[\{V_{\alpha}\}]$ is above a threshold when all potentials are identical, $\lambda$ is determined for which the first non-zero $v_i^{(\alpha)}-v_i^{(\beta)}$ appear(s). Let us denote the subsets of $\alpha$ for which the potentials are different as $A_1$, $A_2$, ..., $A_q$. Now we change the penalty term as follows:
     \begin{equation}
       ||\Delta v||_1 = \sum_{m=1}^q\sum_i\sum_{\alpha < \beta; \alpha,\beta\in A_m}|v_i^{(\alpha)}-v_i^{(\beta)}|\, ,
       \label{eq:norm2}
    \end{equation}
     i.e., we do not penalize anymore differences between potentials that we already identified as different in at least one parameter. Next, if $F[\{V_{\alpha}\}]$ is still above the threshold, we again determine $\lambda$ for which the first non-zero terms (or one term) appear in Eq.~\ref{eq:norm2}, and identify new subsets $A_m$. The procedure is repeated until $F[\{V_{\alpha}\}]$ is below the threshold. The information needed to restore the SOAOs is the set of different vectors ${\bf v}^{(\alpha)}$ and a set of quantum numbers ($n$, $l$) for each of the potentials for the $N_{\eta}$ radial basis functions.

     A complete spline basis set for the same logarithmic radial grid as for SOAOs can be used as the basis $f_i(r)$ for the potentials $V_{\alpha}(r)$ = $\sum_i v_i^{(\alpha)}f_i(r)$. However, for physical reasons $rV_{\alpha}(r)$ are expected to be smooth functions, so that further storage reduction can be achieved by reducing the number of grid points for representing the potentials. Several different grid densities can be tested automatically by repeating the procedure for finding $V_{\alpha}^{\rm opt}$ for each grid and checking if the storage size is reduced. In the worst case scenario, when all $V_{\alpha}$ are different, the required storage may still be reduced, if a sparser grid is found for the potentials without sacrificing the wavefunction representation accuracy.
     
The amount of data for WF storage is significantly reduced, if the WFs themselves are localized. The localization introduces sparsity into the matrix of expansion coefficients in equation~\ref{eq:basisrep} when the WFs are represented in a localized basis (which is the case in our representation), in addition to the reduction in the matrix' size due to the optimization of the basis functions as described above. Since physical quantities remain unchanged upon unitary transformations of the Hilbert space defined by the occupied and unoccupied orbitals, we can find a transformation that maximally localizes the initial set of WFs. The obtained states are called maximally-localized Wannier functions (MLWF)~\cite{marzari2012maximally}. An efficient approach for obtaining MLWF based on compressed sensing was developed~\cite{ozolicnvs2013compressed,ozolicnvs2014compressed}. The approach avoids starting-point dependent non-convex minimization and arbitrary cut-off parameters. Recently, it has been extended to improve efficiency by using a local orthogonality constraint imposed on the Bloch functions in reciprocal space, and to exactly preserve physical symmetries of the system~\cite{budich2014search}. Partially occupied bands (such as conduction bands in metals and semiconductors) require ``disentangling'' before the MLWF can be obtained, as implemented, e.g., in the Wannier90 program~\cite{mostofi2014wannier90}. The resulting MLWFs are approximate, but the accuracy of the transformation is controlled by a well-defined parameter (an energy window), and a very high accuracy can be achieved in practice with minimal additional effort. The gains due to the localization of the WFs increase with the number of atoms and unit cell size, because the number of non-overlapping WFs, and, consequently, the sparsity of the matrix in equation~\ref{eq:basisrep} are also increased. For example, in wide band gap oxides the O 2$p$ valence states are usually delocalized over the entire unit cell. The storage of each O 2$p$ WF represented in a minimal atomic basis in this case would in general require $N_{atoms}$ double-precision numbers, where $N_{atoms}$ is the number of O atoms in the unit cell. However, if the corresponding O 2$p$ Wannier functons can be localized on single O atoms, the amount of data to be stored will be reduced by a factor of $N_{atoms}$.

  Once SOAOs are found, the representation of WFs in terms of the new basis set is obtained by minimizing the difference between the WFs $\psi_{i{\bf k}}^{\sigma}$ represented in the original basis and in the SOAO basis \{$\phi_{\alpha}$\}:
  
  \begin{equation}
    \frac{\partial}{\partial C_{i{\bf k}}^{\sigma \alpha}}\sum_{i{\bf k}\sigma} \int \left( \psi_{i{\bf k}}^{\sigma}({\bf r})-\sum_{\alpha}C_{i{\bf k}}^{\sigma \alpha} \phi_{\alpha}({\bf r})\right)^2 d^3r = 0\, .
    \label{eq:newcoeffs}
  \end{equation}

   The procedure for finding the most compact WF subspace representation described here contains well-defined parameters controlling accuracy of the representation. However, additional constraints can be imposed on the minimization using the method of Lagrange multipliers to avoid errors in important properties of the original WF representation. In particular, the WFs represented in the new basis can be required to be strictly orthonormal, and the electron density to integrate to the number of electrons exactly. With these and other practically relevant constraints, the minimization translates into solving a linear eigenvalue problem.

  In principle, the electron density can be stored as the density matrix in the same basis as WFs, $D_{\alpha\beta}^{\sigma}=\sum_{i{\bf k}}f_{i{\bf k}}^{\sigma}C_{i{\bf k}}^{\sigma \alpha *}C_{i{\bf k}}^{\sigma \beta}$, provided {\em all} occupied states have been used to determine SOAOs. This is equivalent to expanding the density $n^{\sigma}({\bf r})$ in the basis of products of the atomic functions. The most compact representation is achieved when SOAOs are natural atomic orbitals. The density matrix is sparse due to locality of the basis functions ($D_{\alpha\beta}^{\sigma}$ are non-zero only if $\phi_{\alpha}({\bf r})$ and $\phi_{\beta}({\bf r})$ overlap).

  However, apart from the density, it is desirable to store other scalar fields, in particular Kohn-Sham effective potentials and electrostatic potentials. A method to obtain an accurate decomposition of an arbitrary scalar field in terms of atom-centered functions has been developed previously~\cite{delley1990all,blum2009ab}. The method is based on atom-centered ``partition of unity'':
  \begin{equation}
    \sum_Ap_A({\bf r})=1\, ,
  \end{equation}
  where
    \begin{equation}
    p_A({\bf r})=\frac{g_A({\bf r})}{\sum_{A'}g_{A'}({\bf r})}\, ,
  \end{equation}
    with functions $g_A({\bf r})$ strongly peaked and centered at corresponding atom $A$, and the index $A'$ running over all atoms in the unit cell. A scalar field $f({\bf r})$ can then be decomposed as a sum of atom-centered parts:
    \begin{equation}
f({\bf r})=\sum_{A,lm}\tilde{f}_{A,lm}(|{\bf r}-{\bf R}_A|)Y_{lm}(\Omega _A)\, ,
    \end{equation}
    where $Y_{lm}(\Omega _A)$ are spherical harmonics centered at atom $A$, and
      \begin{equation}
        \tilde{f}_{A,lm}(r) = \int_{r=|{\bf r}-{\bf R}_A|}p_A({\bf r})f({\bf r})Y_{lm}(\Omega_A)d\Omega_A\, .
        \label{eq:atompart}
      \end{equation}
      The atom-centered parts are calculated on the radial grids of corresponding atoms, and then interpolated using splines. The integrals in equation~\ref{eq:atompart} can be calculated analytically or numerically using optimized angular girds~\cite{delley1990all,blum2009ab}. While arbitrary atom-centered functions $g_A({\bf r})$ can be used for the ``partition of unity'', the accuracy of the representation may depend on this choice. We suggest to use the partition scheme developed by Stratmann {\em et al.}~\cite{stratmann1996achieving}, since we find it to give accurate results in ES calculations for a wide range of systems~\cite{blum2009ab}. The scalar field $f({\bf r})$ is stored as the values of $\tilde{f}_{A,lm}(r)$ on the radial grid for each atom $A$ and each $(l,m)$ channel, and can be restored on demand as the sum of splined radial functions multiplied by spherical harmonics. The maximum angular momentum $l_{max}$ for the spherical harmonics can be selected based on the convergence of the representation accuracy with $l_{max}$. Functions $\tilde{f}_{A,lm}(r)$ should be stored only if they significantly deviate from zero. Taking the number of radial grid points per atom 150, and $l_{max}$=8 as a safe (upper) limit, we estimate the storage size for a single scalar field to be $\sim$95~kB per atom within unit cell.

\section{Conclusions}
In this work, we present a concept of a code-independent compact representation of one-electron WFs and other volumetric data (electron density, electrostatic potential, etc.) produced by ES calculations. The compactness of the representation insures minimization of digital storage requirements for the computational data, while the code-independence makes the data ready for analysis by artificial intelligence. The implementation of the proposed WF storage concept requires minimal development in addition to natural atomic orbital and Wannier function analysis tools. These types of ES analysis provide important parameters describing ES (descriptors), and are therefore must-have tools for computational data analytics. Thus, the proposed storage concept perfectly fits any infrastructure that makes the analysis tools readily available on code-independent basis. The data stored in this form can be complementary to code-specific storage formats that are used, for example, for restarting ES calculations.

The proposed approach for compact WF storage includes the following four major steps.
First, an atom-centered basis set is generated such that it accurately
represents the set of one-electron states to be stored. This step is only necessary if the WFs are not originally represented in an atom-centered basis (as, e.g., in plane-wave calculations). Second, a minimal set of numeric atom-centered orbitals is found that represents the set of WFs with a desired accuracy. For this, a procedure similar to finding natural atomic orbitals is employed. Third, a set of atom-centered spherically symmetric potentials is determined so that the minimal NAO basis can be calculated numerically on demand by solving the Schr\"odinger equation for a set of orbital quantum numbers. Finally, localized Wannier functions are calculated to introduce sparsity into the expansion of WFs in the minimal NAO basis set.

For scalar fields such as electron density, Kohn-Sham potentials, and electrostatic potentials we propose to use the approach based on atom-centered partition of unity. This approach is used extensively in electronic-structure calculations based on NAOs, and have been demonstrated to be accurate and efficient.

Each step of the above procedure is controlled by well-defined parameters that can be tuned to insure the accuracy of the WF representation within a desired threshold, theoretically achieving an arbitrarily high accuracy, i.e., an information-lossless conversion, albeit at the cost of increased storage needs. Additional constraints can be imposed to avoid errors in important properties of the original WF representation. In particular, the WFs represented in the new basis can be required to be strictly orthonormal, and the electron density to integrate to the number of electrons exactly. A detailed study of the numerical sensitivity of the procedure to the parameters, and of the performance in terms of storage reduction for real-life applications will be discussed in detail in a future publication.

\section{Acknowledgements}
  The work is supported by European Union’s Horizon 2020 research and innovation programme under grant agreement No 676580 via the cluster-of-excellence ``Novel Materials Discovery (NOMAD) Laboratory'' (\url{https://nomad-coe.eu/}). We thank Georg Kresse and Peter Blaha for fruitful discussions. SVL also acknowledges support by the Ministry of Education and Science of the Russian Federation (Grant No. 14.Y26.31.0005) for the development of wavefunction analysis methods, and by the Ministry of Education and Science of the Russian Federation in the framework of Increase Competitiveness Program of NUST MISIS (No K2-2017-080) implemented by a governmental decree dated 16 March 2013, No 211, for the development of tools for materials databases.
      
\section*{References}

\bibliography{references}

\end{document}